\documentclass[11pt,a4paper]{article}
\usepackage{jheppub}

\usepackage{color}
\usepackage{amsmath}

\usepackage{verbatim}   
\usepackage{subfigure}  

\usepackage{amsfonts}
\usepackage{amssymb}
\usepackage{mathrsfs}
\usepackage{graphicx}
\usepackage{slashed}

\newcommand{\kahler}{K\"{a}hler }

  \newcommand{\GeV}{{\mathrm {GeV}}}
   \newcommand{\TeV}{{\mathrm {TeV}}}

\def\inv{^{\raise.15ex\hbox{${\scriptscriptstyle -}$}\kern-.05em 1}}
\def\lbar{{\lower.35ex\hbox{$\mathchar'26$}\mkern-10mu\lambda}} %lambda bar

\def\to{\rightarrow}

\let\<=\langle
\let\>=\rangle

\let\+=\uparrow

\let\l=\left
\let\r=\right
\begin{document}

\hfill \vspace{-5mm} OUTP-13-03P

%%%%%%%%%%%%%%%%%%%%%%%%%%%%%%%%%%%%%%%%%%%%%%%%%%%%
%%%%%%%%%%%%%%%%%%%%%%%%%%%%%%%%%%%%%%%%%%%%%%%%%%%%%

\title{Retrofitted Natural Supersymmetry from a U(1)}

\author[a]{Edward Hardy,}
\emailAdd{e.hardy12@physics.ox.ac.uk}
\author[a,b]{John March-Russell}
\emailAdd{jmr@thphys.ox.ac.uk}
\affiliation[a]{Rudolf Peierls Centre for Theoretical Physics,
University of Oxford,\\
1 Keble Road, Oxford,
OX1 3NP, UK}
\affiliation[b]{Stanford Institute for Theoretical Physics, Department of Physics,\\
 Stanford University, Stanford, CA 94305, USA}
%

%%
%
%\date{\today}
%
%

\abstract{We propose that a single, spontaneously broken, U(1) gauge symmetry may be responsible for suppressing both the first two generation Yukawa couplings, and also, in a correlated manner, parameters in the dynamical supersymmetry (SUSY) breaking sector by the mechanism of retrofitting. In the dynamical SUSY breaking sector, these small parameters are typically required in order to introduce R-symmetry breaking in a controlled manner and obtain phenomenologically viable meta-stable vacua.  The heavy U(1) multiplet mediates a dominant contribution to the first two generation MSSM sfermion soft masses, while gauge mediation provides a parametrically suppressed soft term contribution to the stop and most other states, so realising a natural SUSY spectrum in a fashion consistent with SUSY unification.  In explicit models the spectra obtained can be such that current LHC limits are evaded, and predictions of flavour changing processes are consistent with observation.  We examine both implementations with low scale mediation, and string-motivated examples where the U(1) is anomalous before the inclusion of a generalised Green-Schwarz mechanism.

%XXXXWe study models where the same small parameter leads to Froggatt-Nielsen style suppression of the first two generation MSSM fermion masses, and also acts to provide a small parameter in a dynamical supersymmetry breaking sector in the style of retrofitting. In cases where this suppression is generated as a result of an underlying gauge symmetry, a contact operator from the heavy gauge multiplet leads to an additional contribution to supersymmetry mediation. We examine a purely field theory model realised at a low scale and also a string motivated case where the small parameter is generated by an anomalous global symmetry that is the remnant of a gauge symmetry broken by the Stueckelberg mechanism. Models of this type are found to straightforwardly realise natural supersymmetric spectra in a phenomenologically viable manner with the first two generation squarks beyond current LHC reach but not so heavy as to drive the stops tachyonic during running.
}

\maketitle

\section{Introduction}

If softly-broken supersymmetry (SUSY) is to be a successful theory of the weak-scale, solving the hierarchy problem, then it must
meet a number of serious challenges.  First, on the theoretical side, there is still significant uncertainty over the mechanism of SUSY breaking and its mediation to the visible sector.   From the perspective of the hierarchy problem the most attractive possibility, as first argued by Witten \cite{Witten:1981nf}, is the dynamical breaking of SUSY via dimensional transmutation and non-perturbative effects since this can occur at an energy scale which is exponentially separated from other scales in the theory, hence naturally leading to small soft terms in the visible sector.  Despite this attractive feature, many models of dynamical supersymmetry breaking (DSB) still require small parameters, or masses to be parametrically suppressed relative to other scales in the theory.  Well known examples include the 3-2 model \cite{Affleck1985557} and, especially, the ISS model \cite{Intriligator:2006dd}  where small parameters are required to ensure that (in the presence of the phenomenologically required R-symmetry breaking) a possibly metastable vacuum is sufficiently long-lived to be viable.

An appealing approach to deal with this is through so-called \emph{retrofitting} \cite{Dine:2006gm}, where IR irrelevant operators generate small parameters which would otherwise be forbidden by symmetries of the theory. In the case that the  operators introduce a small amount of R-symmetry breaking, this is not a surprising scenario, as vanishing of the vacuum energy post SUSY breaking requires the superpotential to have a R-symmetry violating expectation value, 
which is transmitted through supergravity to produce the required operators.

Second, on the phenomenological side, there is increasing tension between the requirement that superpartners should be close to the electroweak scale to prevent the reintroduction of a little hierarchy problem and negative results of collider searches first at LEP and now at the LHC \cite{Beringer:1900zz}.  One possibility to weaken these experimental constraints is to make the first two generation sfermions somewhat heavy, while keeping third-generation squarks, especially stops, and electroweak gauginos and higgsinos light, the so called `Natural SUSY' scenario \cite{Dimopoulos:1995mi,Cohen:1996vb}.  This suppresses production of sparticles at the LHC while reducing (though not completely eliminating) the fine tuning of the electroweak scale.  Although there has been much phenomenological study of this case (see for example \cite{Papucci:2011wy}) it is unclear how such spectra may be realised from a UV theory in a way that maintains the successes of low-energy SUSY such as the gauge unification prediction of $\sin^2\theta_w(M_z)$ and radiative electroweak symmetry breaking.   The issue is again the appearance of small parameters, in this case the ratio of third to first and second generation sfermion masses.   Of course, there is another well known problem involving unexpectedly small parameters: the Standard Model fermion masses themselves which exhibit a large range of values.  A very popular way of dealing with this, which again involves irrelevant operators generating terms that are forbidden at leading order by a new symmetry, is the Froggatt-Nielsen mechanism \cite{Froggatt:1978nt}.

%%%%%%%%%%%%%%%%%%%%%%%%%%%%%%%%%%%%%%%%%%
Our aim in this work is to argue that these disparate cases are in fact directly related, with the same broken symmetry leading to small parameters in both the SUSY-breaking and visible fermion/sfermion flavour sectors. In particular we consider theories where there is an additional underlying U(1) gauge symmetry broken at high scale.  While such symmetries may simply be regarded as a feature of an effective theory, they often automatically appear in an underlying string theory model. The later case is highly attractive since these symmetries are naturally anomalous in the field-theory limit before a generalised Green-Schwarz mechanism is included which typically leads them to gain a large SUSY preserving mass \cite{Ibanez:1999it}. Additionally, they can be linked to generation of the visible sector fermion masses in brane stack models, whereby different generation fermions are charged differently, as recently discussed in \cite{Berenstein:2010ta,Berenstein:2012eg}.

Of course the possible role of U(1) gauge symmetries in breaking SUSY and mediating this to the MSSM sector has been studied extensively and it has previously been proposed to generate flavour structure in sfermion masses, see e.g., \cite{Dvali:1996rj,Nelson:1997bt,Mohapatra:1996in,Hisano:2000wy,Raby:1998bg,Brax:2000ip,Eyal:1999hd,Badziak:2012rf,Komine:2000jr,Faraggi:1997be,Babu:2005ui,Aharony:2010ch,Dudas:2008qf,Choi:2006bh,Acharya:2011kz}.\footnote{Alternative UV models that could realise natural spectra have also been proposed \cite{Nomura:2007ap,Craig:2012yd}, and similar spectra with light stops and the other sfermions heavy may occur in M-theory compactifications after running \cite{Acharya:2008zi}.}  Many previous models have proposed the fields that break the U(1) can be directly involved in the SUSY breaking sector. While this is an attractive prospect it leads to issues such as the dilaton necessarily gaining an F-term that may dominate the mediation \cite{ArkaniHamed:1998nu}.  An additional problem is if gaugino and third generation soft masses are generated through gravity mediation it is very hard to avoid dangerous flavour changing processes without making the first two squarks generations so heavy as to drive the stops tachyonic during running \cite{ArkaniHamed:1997ab}.

In contrast, in the models we consider, the U(1) vector multiplet gains a mass at a high scale and only acts as an additional messenger interaction without being directly involved with the SUSY breaking.  Importantly, since the SUSY breaking sector is charged under this gauge symmetry, there is an additional contribution to MSSM soft masses from a contact interaction after integrating out the heavy vector multiplet.\footnote{Operators generated by integrating out heavy gauge fields have previously been proposed as a viable mechanism of mediating supersymmetry breaking  \cite{Nardecchia:2009nh,Nardecchia:2009ew}, and have been studied in the context of dynamical SUSY breaking and gauge mediation with universally charged MSSM fields \cite{Caracciolo:2012de}.} Then, as we will argue, if only the first two generation sfermions are charged under the broken U(1), this can lead to first two generation sfermion soft masses a factor of a few larger than the gauge mediated soft masses, and therefore the first two generation sfermions can be heavy enough to evade detection and realise natural SUSY, but not so heavy as to drive the stop tachyonic through RG running.  However, with first two generation sfermions in the mass range $\sim {\rm few}$~TeV, flavour violation is not adequately suppressed unless there is a high degree of degeneracy between these sfermions.  Because of this we take the first two generations to be charged equally under the U(1), so both broken U(1) mediation and the competing SM gauge mediation are flavour universal, leading to flavour observables within current limits.  Of course, a consequence of this is that the observed hierarchies in first and second generation fermion masses and mixing cannot be `explained' by selection rules following from the breaking of U(1), and only the hierarchy and mixing between the third generation and the lower generations is due to the Froggatt-Nielsen mechanism.  Our attitude here is that the flavour structure of the first two generations is set by high scale physics which is independent of SUSY breaking dynamics.   As we will show this is allowed since, in our model there can be ${\cal O}(1)$ breaking of the flavour symmetry of the lower generation fermions consistent with the fact that the sfermion partners simultaneously possess an effective
flavour symmetry that is only very weakly broken at loop order by the tiny lower-generation Yukawas.

Turning to the organisation of our paper, in Section~\ref{field} we introduce the overall structure of our models and the basic mechanisms of SUSY breaking and mediation in a field theory setting, illustrating the ideas first using a Polonyi model, and then a fully dynamical ISS model. Following this in Section~\ref{string}, we examine how such models may naturally appear from an underlying string theory possessing anomalous U(1) gauge symmetries. In Section~\ref{soft} we consider the low energy spectrum of soft terms obtained, while in
Section~\ref{pheno} we note some additional interesting phenomenological possibilities and discuss signatures.  Finally, two Appendices contain technical details.
 
%%%%%%%%%%%%%%%%%%%%%%%%%%%%%%%%%%%%%%%%%%%%%%%%%%%%%
\section{Structure of Field Theory Implementation} \label{field}
We begin by discussing the implementation of our models in a low energy field theory setting using a Polonyi model as a straightforward example of the SUSY breaking sector. Following this we implement a fully dynamical example, an ISS model.

\subsection{Low Energy Polonyi Model} \label{pol}
The underlying theory is specified by four sectors. At the highest scale a sector that breaks the U(1), a DSB sector, a messenger sector, and the visible sector. All sectors involve fields charged under the U(1) symmetry, and the superpotential takes the form
\begin{equation}
W = W_{U\left(1\right)}+W_{DS} + W_{mess}+ W_{visible} ,
\end{equation}
with a UV cutoff at a scale $M_*$ and canonical \kahler potential up to irrelevant operators suppressed by powers of $M_*$. The sector $W_{U\left(1\right)}$ involves only fields $S_i$, with U(1) charge i, and spontaneously breaks the U(1) gauge symmetry through fields $S_{1}$ and $S_{-1}$ gaining vacuum expectation values (VEVs) $v$ leading to the gauge boson gaining a mass $M_{Z'} = g' v$. Here $g'$ is the U(1) gauge coupling, and we assume that the scale of $ W_{U\left(1\right)}$ is sufficiently above the other sectors that the VEVs of fields $S$ are rigidly fixed. Hence, once this symmetry breaking occurs, the fields $S_{\pm 1}$ in the other sectors may be replaced by their expectation values $v$. This leads to a small ratio in the theory we denote by $\epsilon = \frac{v}{M_*}$.

The DSB sector has fields charged under the U(1) symmetry and the superpotential includes irrelevant operators generated at the cutoff of the theory with the form
\begin{equation}
\Delta W_{DS} = \frac{S^n}{M_*^m} \mathcal{O}_{DS}
\end{equation}
where $n$ and $m$ are integers and $ \mathcal{O}_{DS}$ are operators involving the fields in this sector. Once some of the $S_i$ gain a VEV these couplings lead to small mass terms and parameters. In particular, consider a very simple sector of Polonyi form with one field $\Phi$ with charge $+6$ under the U(1) symmetry.\footnote{In a realistic model this would typically be a composite field, hence a relatively large charge assignment is not especially unusual.} After $S_{\pm 1}$ gain their common VEV, the superpotential is 
\begin{equation}
W\supset \frac{S_{-1}^{6}}{M_*^4}\Phi = \epsilon^4 v^2 \Phi ,
\end{equation}
leading to a SUSY-breaking F-term $F_{\Phi} = \epsilon^4 v^2$.

In the messenger sector there are fields, $\{ \psi, \psi^c\}$, that form a vector-like pair under the SM gauge groups which act as messengers of gauge mediation. They are charged under the U(1) with couplings to the fields $S$ and also to the DSB sector of the form
\begin{equation}
 W_{mess} = \frac{S^{p+1}}{M_*^p} \psi \psi^c + \mathcal{O}_{DS} \psi \psi^c .
\end{equation}
We further assume they have a potential (either at tree or loop level) such that the SUSY breaking minimum remains either a stable or metastable state. Taking the combination $\psi \psi^c$ to have charge  $-4$ this sector includes a mass term and interactions with $\Phi$ given by
\begin{equation}
\begin{aligned} \label{gaugecoupling}
W&\supset \frac{S_{-1}^2}{M_*^2} \Phi \psi \psi^c + \frac{S_{1}^{3}}{M_*^4} \psi \psi^c \\
&=  \epsilon^2  \Phi \psi \psi^c + \epsilon^3 v\psi \psi^c .
\end{aligned}
\end{equation}
Due to the coupling between the field $\Phi$ and the messengers, there will be gauge mediated soft masses roughly of size
\begin{equation}
\label{gauge}
m_{\rm{gauge}}\sim\left(\frac{\alpha}{4\pi}\right) \frac{F_{\rm{eff}}}{M_{\rm{mess}}} \sim \left(\frac{\alpha}{4\pi}\right) \epsilon^3 v,
\end{equation}
where $F_{\rm{eff}}= \epsilon^6 v^2$ is the effective F-term felt by the messenger fields due to its coupling to $\Phi$. In order that these soft terms are close to the electroweak scale, for values of $\epsilon$ appropriate to fermion masses,  $v$ and $M_*$ must be relatively close to the weak scale, hence this is very low scale gauge mediation with messenger masses an inverse loop factor above the weak scale. This is phenomenologically beneficial as it results in relatively little running and the first two generations can be pushed heavier without leading to a tachyonic stop.

Finally the visible sector superpotential takes the form
\begin{equation}
W_{visible} = c_{ij} \left(\frac{S_{-1}}{M_*}\right)^{q_{ij}}  \mathcal{O}_{visible}^{ij} ,
\end{equation}
where $\{i,j\}$ are generation indices.
The parameters $c_{ij}$, which are not constrained by the U(1) symmetry, are set by UV physics at (or above) the scale
$M_*$ where the irrelevant operators are generated, and may or may not satisfy other symmetry relations.  After U(1) symmetry
breaking the effective Yukawa couplings relevant to IR physics which set the observed fermion mass ratios and CKM mixings
are 
\begin{equation}
\lambda_{ij} = \epsilon^{q_{ij}} c_{ij}. 
\end{equation}
As is well known, the observed 3rd-generation fermion masses and mixings have properties which set
them apart from the lower generations: Not only is the top Yukawa coupling ${\cal O}(1)$ (as can be those of the bottom and tau if $\tan\beta$ is large)
in distinction to the suppressed lower-generation couplings, but SU(5) SUSY unification predictions work well for $m_b/m_\tau$ while the analogous
lower-generation predictions fail badly. In addition, if the experimentally observed ratios of second generation to third
generation fermion masses at a low scale are run to the GUT scale, assuming weak scale SUSY, the resulting ratios and mixings
$m_c/m_t \approx 1/300$, $m_s/m_b \approx 1/40$, $m_\mu/m_\tau \approx 1/17$ and $V_{cb} \approx 1/25$ are well-described
by a structure of Yukawa couplings for the up and down quarks and leptons depending on a single small parameter $\epsilon\sim1/20$
of the form
\begin{equation}
U \simeq \left(\begin{array}{cc}
 \epsilon^2 &\epsilon\\
 \epsilon & 1
\end{array}\right),\qquad
D \simeq \left(\begin{array}{cc}
 \epsilon &*\\
** & 1
\end{array}\right),\qquad
E \simeq \left(\begin{array}{cc}
 \epsilon &*\\
** & 1
\end{array}\right),
\end{equation}
where here ``$*$" and ``$**$" denote entries that are ${\cal O}(1)$ (respectively ${\cal O}(\epsilon)$) or smaller, see e.g. \cite{Hall:2001rz,Hebecker:2002re}.  This
structure strongly suggests that some dynamics sets this pattern, such as that following from a Froggatt-Nielsen mechanism \cite{Froggatt:1978nt}, or
from extra-dimensional orbifold-GUT constructions \cite{Hall:2001pg,Hebecker:2001wq,Hall:2001rz,Hebecker:2002re}. 
This is particularly the case since, as far as we are aware, there is no anthropic reason for the second-generation masses and $V_{cb}$ to take their observed values.
On the other hand the masses of the first generation quarks, as well as the mass of the electron, do not fit so nicely with any simply dynamical mechanism depending on only
one small parameter, and are, in addition, (remarkably) in accord with the anthropic ``catastrophic boundaries'' linking $m_u$, $m_d$, $m_e$, with $\lambda_{QCD}$ and $\alpha_{em}$
\cite{Hogan:1999wh,Hall:2007ja}.  

Because of this we now make the crucial assumption, different from many previous studies, that the physics that sets the 2-3 inter-generational mass ratios
and mixings is different than that which sets the 1-2 ratios and mixings. Specifically our starting place is that second-third generational physics
is set by the U(1)-dependent factors $\epsilon^{q_{ij}}$ while the first-second generation physics is set by the $c_{ij}$'s which are {\it not}
determined by our broken gauged U(1).\footnote{The mixings and hierarchies between the lighter two generations may result from another broken Froggatt-Nielsen
flavour symmetry such as U(1) or SU(2) which is either not gauged, or does not interact with the SUSY breaking sector, or, alternatively may instead be the result
of landscape scanning of the coefficients $c_{ij}$ subject to the strong anthropic constraints that they must obey.  The important point for our work is that we do not need
to specify this physics as long as it is independent of (commutes with) our U(1) that interacts with the DSB and similarly retrofits some if its couplings.}

In detail, the up-like-Higgs and top quark multiplets are uncharged under the U(1), such that a superpotential term $W\supset H_{u}q_{L3}u^{c}$
is allowed, and is expected to have an order 1 coefficient, as observed. In contrast first two generation fields of the same SM quantum numbers are taken to be charged {\it equally} under
the U(1), leading to mass terms that are suppressed by equal powers of $\epsilon$. Since the U(1) symmetry is abelian the ${\cal O}(1)$ coefficients that dress these couplings possess no symmetry properties, and can lead to the observed mass splitting of the first two generation visible sector fermions and the Cabbibo mixing structure, as we discuss in Section~\ref{soft}. (Later, in Section \ref{soft} when investigating particular models, we give explicit charge assignments and show the textures generated in the visible fermion masses.) 

For phenomenologically viable charge assignments, including only MSSM matter, the U(1) symmetry would appear to have anomalies of the form $\rm{U(1)}\times G_{visible}^2$ and $\rm{U(1)}^2 \rm{U(1)}_Y$, however these can be cancelled by the messenger fields (or other matter which is chiral under the U(1) and vector like under the visible sector groups). Choosing a GUT compatible U(1) charge assignment for the visible sector allows these anomalies to be cancelled by matter in complete GUT multiplets hence gauge unification is preserved.\footnote{It is possible to arrange, either by choice of U(1) charges or by the geometric localisation of the messenger fields, that less suppressed interactions between $S$ and the visible sector are not generated upon integrating out these states.}

After U(1) symmetry breaking, as a result of integrating out the heavy U(1) gauge boson there will be a \kahler contact operator \cite{ArkaniHamed:1998nu}, derived in Appendix \ref{apenA}, between any two fields charged under the U(1) symmetry. This is important for our phenomenology as it leads to an extra coupling between the field which gains an F-term, $\Phi$, and the first two generation MSSM fields (and third generation down type quarks and leptons), $Q_{1,2}$,
\begin{equation}
\int d^4\theta c_i g^2 \left(\frac{\Phi^{\dagger}\Phi Q_{1,2}^{\dagger}Q_{1,2} }{M_{Z'}^2}\right) .
\label{kah}
\end{equation}
Here $M_{Z'}$ is the mass of the heavy U(1) gauge boson, while $c_i\sim q_\Phi q_{1,2}$ depends on the U(1) charges
of the fields. Since $M_{Z'}= g' v$ the dependence on $g'$ drops out leading to soft masses for the first two generations
\begin{align}
m_{K}^2 &= - c_i \frac{\left|F_{\Phi}\right|^{2}}{v^2} . \label{kahmass}
\end{align}
At the scale these interactions are generated, the coefficients $c_i$ depend only on the U(1) gauge charges of the fields, and therefore can naturally be equal for the first two
generations by a discrete choice of the charges. During RG evolution down to the weak scale the dominant running effects will be due to SM gauge interactions which
are still universal.  The only deviations from universality are due to the first and second generation Yukawas and have a negligible effect. Therefore, flavour changing currents are not generated in the visible sector, which is crucial for acceptable spectra. In order that sfermions gain a positive mass contribution,
it must be assumed that $c_{1,2}<0$ . In Appendix \ref{apenA} it is seen that this  can be easily realised in UV completions, simply if the fields $\Phi$ and $Q_{1,2}$ have the same sign charge. Since the exact properties of the underlying theory are unknown, we fix and overall normalisation by setting $c_{i}=-q_i$ for all states in the visible sector with U(1) charge
$q_i$. The qualitative properties of the spectra obtained are not especially sensitive to this assumption.

Since the third generation up type quarks are uncharged under the U(1) there are no such terms generated for the stops through this type of interaction. Further, integrating out the gauge multiplet will not generate terms of the form $\int d^{4}\theta f\left(\Phi^{\dagger},\Phi\right)Q_{1,2}^{\dagger}Q_{3}$ hence these are suppressed relative to the \kahler mass terms \eqref{kahmass}. Since the Higgs fields are uncharged under this symmetry, the soft masses $M_{H_d}^2$ and $m_{H_u}^2$ are not large which is beneficial in avoiding large fine tuning of the electroweak scale. An important assumption we are making of the UV theory is that there is no additional field content that generates significant \kahler couplings between $\Phi$ and the top multiplet. Of course in a realistic UV completion these will appear at some level, however may naturally be expected to be suppressed by either $M_*$ or $M_{pl}$ and therefore be negligible compared to the other contributions.

In the model considered here the interaction \eqref{kahmass} generates masses
\begin{align}
m_{K}^2 &= - c_i \frac{\left|F_{\Phi}\right|^{2}}{v^2} = \epsilon^8 v^2 .
\end{align}
There will be a similar coupling to the messenger fields,
\begin{equation}
\int d^4\theta c_i \frac{\Phi^{\dagger}\Phi}{v^2}\psi^{\dagger}\psi .
\end{equation}
Since the messenger mass is close to $\sqrt{F}$, the SUSY breaking from this term can lead to slight corrections to the gauge mediated masses induced in the visible sector compared to the normal formula derived assuming analytic continuation. In Appendix \ref{apenB} we give the general formula which we use in our later phenomenological analysis. While the exact form of these corrections is complicated their effect is straightforward: both the next corrections in $\frac{F}{m}$ and those from SUSY breaking diagonal masses tend to increase the sfermion masses relative to gaugino masses as they do not break R-symmetry.

The key phenomenological feature of our models is that the ordinary gauge mediated contribution from messenger fields will compete with this \kahler contribution to the first two generations leading to a scenario where the first two generation sfermions become relatively heavy, while the stop quarks stay light, realising a natural SUSY spectrum. These contributions can give phenomenologically reasonable soft terms and natural spectra with appropriate choices of $M_*$ for $\epsilon$ in the range suggested by fermion mass hierachy. In Section~\ref{soft} we study the MSSM spectra for reasonable choices, however first we examine a more sophisticated and UV complete model of SUSY breaking.

%%%%%%%%%%%%%%%%%%%%%%%%%%%%%%%%%%%%%%%%%%%%%%%%%%%%%%%%%%%%%%
%%%%%%%%%%%%%%%%%%%%%%%%%%%%%%%%%%%%%%%%%%%%%%%%%%%%%%%%%%%%%%

\subsection{ISS Model of Supersymmetry Breaking and Mediation}
\label{softiss}
While general features of these models can be realised in many examples we now consider the ISS model \cite{Intriligator:2006dd} as an example of a fully dynamical SUSY breaking sector, which, once including the suppression from retrofitted couplings, needs no small scales or couplings. In particular this allows the very natural possibility of associating $M_*$ with the GUT scale. A retrofitted model has previously been studied in \cite{Brummer:2007ns}, and here we consider a mediation to the visible sector through the addition of messenger fields.

Consider supersymmetric QCD with gauge group $SU\left(N_c\right)$ and $N_f$ quarks, $Q_i$, $\tilde{Q_i}$, in the range $N_c+1\leq N_f < \frac{3}{2}N_c$. The quarks have charge $+n/2$ and the messenger fields $-n/2$ under the U(1) which is broken in a separate sector by two fields with charge $\pm 1$, $\left<S_{\pm1}\right>=v_{}$.
The $SU\left(N_c\right)$ gauge coupling is asymptotically free and the theory has a dynamical scale, $\Lambda$, above which the superpotential is given by
\begin{equation}
W = \frac{1}{M_*} Q_i \tilde{Q}^i \psi \psi^c + \frac{S_{-1}^{n}}{M_*^{n-1}} Q_i \tilde{Q}^i +  \frac{S_{1}^{n}}{M_*^{n-1}} \psi \psi^c  .
\end{equation}

Below $\Lambda$ the theory is given by the Seiberg dual which consists of magnetic degrees of freedom: dual quarks $q$, $\tilde{q}$ and the meson of the electric theory (canonically normalised following convention) $\Phi_i^j = \frac{Q_i \tilde{Q}^j}{\Lambda}$ with superpotential
\begin{equation}  \label{gaugecouplingiss}
W= \Phi_i^j q^i \tilde{q}_j + \frac{v_{}^{n}}{M_*^{n-1}} \Lambda Tr\left(\Phi\right) + \left( \frac{v_{}^n}{M_*^{n-1}} + \frac{\Lambda}{M_*} Tr\left(\Phi\right) \right) \psi \psi^c .
\end{equation}
With this superpotential, neglecting the small coupling to the messenger fields, the F-terms of the meson field are given by
$F_{\Phi_j^i} = \tilde{q}_j q^i - m \Lambda \delta^i_j$, where $m= \frac{v^n}{M_*^{n-1}}$.

As usual in ISS models the differing ranks of the two contributions to $F_{\Phi}$ imply that not all F-terms can vanish and therefore SUSY appears to be broken. In fact a full analysis using a Coleman-Weinberg potential shows there is a SUSY breaking vacuum at
\begin{equation}
\begin{aligned}
\Phi_j^i &= 0 \\
q^i=\tilde{q}_j^T &=  \left(\begin{array}{c} \sqrt{m \Lambda} \,\, 1_{F-N} \\ 0 \end{array} \right) ,
\end{aligned}
\end{equation}
with $\Phi$ gaining an F-term of order\footnote{This F-term depends on an $\mathcal{O}(1)$ coefficient, which is undetermined by holomorphy and therefore unknown.  However all the soft mass contributions will be seen to depend on $F_{\Phi}$ in the same way therefore this leads to no alteration in the phenomenology.}
$F_{\Phi} \sim \frac{v^n\Lambda}{M_*^{n-1}}$. Since the mass term in the electric theory, $\frac{v^n}{M_*^{n-1}}$, can naturally be much smaller than $\Lambda$, the F-term can be suppressed away from other scales in the theory allowing for small SUSY breaking soft terms to be generated in the visible sector after mediation.

The SUSY breaking vacuum found here is necessarily metastable, as the original theory is vector-like with a non-zero Witten index \cite{Witten:1982df} and therefore must have a SUSY preserving vacua at some point in field space. This occurs at a large VEV of the meson field where the mass deformation term is large and the Seiberg duality is not valid. The lifetime of the SUSY breaking vacua has been estimated \cite{Tamarit:2011ef} and can easily be much longer than the age of the universe. Additionally, it has been argued in \cite{Abel:2006cr}  that in the early universe the theory may be driven towards the metastable SUSY breaking vacuum.

The SUSY breaking in the magnetic theory can be understood in terms of an approximate R-symmetry. Considering the limit where the couplings to the messengers are zero, the theory is the most generic possible with fields charged under an R-symmetry as $\left[ \Phi \right] = 2$ and $\left[q\right] = \left[ \tilde{q} \right] = 0$. Therefore consistent with general theorems \cite{Nelson:1993nf} SUSY is broken. Including the non-renormalisable couplings to messengers in the electric superpotential explicitly breaks this R-symmetry and so the SUSY breaking vacuum is metastable. However the R-breaking is small so it can be long lived. This can be connected to the requirement that the cosmological constant vanishes by making the sector that gives the VEV $\left<S\right> \neq 0 $ the same sector that gives a constant contribution to the supergravity scalar potential.

Regarding the visible sector soft masses gauge mediation will give a contribution
\begin{equation}
m_{\rm{gauge}}\sim\left(\frac{\alpha}{4\pi}\right) \frac{M_*^{n-2}\Lambda F_{\Phi}}{v^n}\sim \left(\frac{\alpha}{4\pi}\right) \frac{\Lambda^2}{M_*} ,
\end{equation}
which can be close to the electroweak scale without fine tuning, as a large hierachy between $\Lambda$ and $M_*$ is natural. In addition
as in the simple Polonyi model, integrating out the heavy U(1) gauge boson leads to a \kahler contact operator between $\Phi$ and other U(1) charged fields.
In the electric theory this is given by
\begin{equation}
\int d^4\theta c_i g^2 \left(\frac{\tilde{Q}^{\dagger} \tilde{Q}+Q^{\dagger}Q}{M_{Z'}^2}\right)Q_{MSSM}^{\dagger}Q_{MSSM} .
\end{equation}
In the magnetic regime we expand $Q = \sqrt{\Phi_0 + \Phi}\sim \Phi_0+ \frac{1}{2}\Phi$ near the origin giving
\begin{equation}
\int d^4\theta c_i g^2 \frac{\Phi^{\dagger}\Phi}{M_{Z'}^2}Q_{MSSM}^{\dagger}Q_{MSSM} . \label{kahiss}
\end{equation}
This induces masses for the first two generations
\begin{equation}
m_K^2 = - c_i \frac{\left|F_{\Phi}\right|^{2}}{v^2} = -c_i \epsilon^{2n-2} \Lambda^2 .
\end{equation}
As in the previous model, there will also be a coupling to the messenger fields:
\begin{equation}
\int d^4\theta c_i \frac{\Phi^{\dagger}\Phi}{v^2}\psi^{\dagger}\psi .
\end{equation}

The qualitative features of such a model are rather similar to that of the simple Polonyi case. Some details differ, however. In particular since the soft terms are set by the dynamical scale $\Lambda$ which can be exponentially separated from $M_*$ (and in fact must be for reasonable gauge mediated soft masses), the two scales $v_{}$ and $M_*$ can now be large.

%%%%%%%%%%%%%%%%%%%%%%%%%%%%%%%%%%%%%%%%%%%%%%%%%%%%%%%%%%%%%%%%%%%%%

\section{String Theory Implementation} \label{string}
Since with phenomenologically viable charge assignments the U(1) symmetry naturally possesses mixed anomalies with the SM gauge groups (at least at the level of triangle diagrams involving chiral fermions, and before including the contribution from messenger fields), it is tempting to associate it with the ``anomalous" symmetries necessarily found in realistic compactifications of string theories which are rendered consistent by the generalised Green-Schwarz mechanism.   While there are various possible stringy UV completions of our models we focus on IIB theories as we now explain.\footnote{Our summary of the appearance of such symmetries in string theory follows the discussion in \cite{Ibanez:1999it} which contains further details.}

In traditional heterotic string theory a U(1) with anomalies cancelled by the Green-Schwarz mechanism necessarily gains a large Fayet-Iliopoulos term $\xi = g^2 M_{pl}^2 \delta_{GS}/ 16\pi^2$ where $ \delta_{GS}$ is the mixed U(1)-gravity$^2$ anomaly coefficient which must be non-zero (though see \cite{Ludeling:2012cu}). Then the D-term contribution to the action is given by $\frac{g^2}{2}( \xi + \sum_{S_j} j S_j K_j )^2$ where $S_j$ are all fields charged under the U(1) (with charge $j$), and $K_j$ is the derivative of the \kahler potential with respect to $S_j$. In order that this does not lead to excessively large SUSY breaking at least one of the fields must gain a VEV, and since this VEV is automatically as large as the mass of the U(1) gauge boson no approximate global symmetry survives in the effective theory below the gauge boson mass.  A theory of this type could in principle be used to generate retrofitted models of the form discussed in the previous Section if the irrelevant operators appear in the effective field theory by integrating out matter of typical mass $M_{pl}$.\footnote{In this case there is the beneficial feature that the ratio $\frac{\left< S_i \right>}{M_{pl}}\sim 0.01$ is automatically appropriate for the fermion mass hierarchies as has been noted by many authors.}  However for our particular case there are some problems with using this traditional heterotic
construction.  In particular, the requirement of universal mixed anomalies (up to Kac-Moody level factors) too-severely restricts our model-building freedom, while the form
of the D-term with non-zero FI term implies that only fields of either positive or negative charge will gain VEVs, not both.  Hence, we consider a slightly different scenario using an underlying IIB string theory (such a IIB construction was recently used to implement a Froggatt-Nielsen mechanism in \cite{Berenstein:2010ta,Berenstein:2012eg}), which leads to a similar but not identical structure to the models of the previous Section. 

In Type IIB string theory, unlike in traditional heterotic theories, {\it non-universal} mixed anomalies can be cancelled by massless twisted
closed string modes which shift under an anomalous transformation.  In the process the U(1) gauge boson gains a mass through the Stueckelberg mechanism.
An important difference with the heterotic case is that, depending on the underlying geometry, the Fayet-Illiopoulos term can be zero, allowing in the IIB case the situation
where no fields charged under the U(1) symmetry necessarily gain VEVs.  Hence, at the perturbative level, a global U(1) symmetry can survive in the low-energy theory
below the mass of the vector boson, this symmetry only being explicitly broken by non-perturbative effects which can naturally be very small \cite{Ibanez:1999it}. The charges of fields under the global U(1) are identical to their charges under the gauged U(1).  One further advantage of IIB models is that by utilising intersecting brane stack constructions it is straightforward to build theories such that only some generations are charged under the anomalous U(1).

With this UV completion, the structure of our models is as follows. At the string scale $M_*= M_{string}$ there is an anomalous U(1) gauge symmetry. Through the Stueckelberg mechanism the associated gauge boson gains a mass $M_{Z'}$ leaving an (approximate, anomalous) global symmetry.  Integrating out this heavy state leads to \kahler 
contact operators with coefficients determined by the charges of the fields involved and the gauge boson mass. Often it is assumed that the vector boson mass is given by $g M_*$. However as shown in \cite{Ghilencea:2002da} this relation can be modified in the case of asymmetric compactifications by ratios of volume factors, which can be parametrically less than 1. We include these effects though a parameter $\lambda$ and write $M_{Z'}=\lambda g M_*$. 

At a lower energy scale the approximate global symmetry is broken by fields $S_{1}$ and $S_{-1}$ gaining common VEVs $v$ with $\epsilon = \frac{v}{M_*}\ll 1$ (these VEVs slightly
correct the vector boson mass).  As in the previous Section, fields in the DSB sector and the visible sector have U(1) charges such that global symmetry forbids some mass terms and parameters at leading order, these terms being generated from irrelevant operators of the form $W_{DS}\supset \frac{S^n}{M_*^m} \mathcal{O}_{DS}$ and  $W_{visible}\supset \frac{S^p}{M_*^p} \mathcal{O}_{visible}$, so suppressing couplings by powers of $\epsilon$.

The resulting soft term structure at the scale of SUSY breaking is similar to the field theory case. There will be a universal gauge mediated contribution and also masses from contact terms generated between fields in the SUSY breaking and visible sectors as a result of integrating out the heavy gauge boson. In the present models $M_{Z'}= g\lambda M_*$, which is of slightly different parametric form compared to the field theory implementation, resulting in a small shift in the relative size of the \kahler contribution. For example, in the ISS model, the \kahler mass contribution to the first two generation sfermions \eqref{kahiss} is
\begin{equation}
m_{K}^2 = - c_i \frac{\left|F_{\Phi}\right|^{2}}{\left(\lambda M_* \right)^2} = -c_i\left(\frac{ \epsilon^{n} \Lambda}{\lambda}\right)^2  .
%\\
%&= -\frac{c_i}{\lambda^2} \left(\frac{v_{}^{n}}{{M_*}^{n}}\right)^2 \Lambda^2 \\
%&.
\end{equation}
One notable change in the phenomenology is that the scale of mediation is typically high. As a consequence, there will be large logs when the soft masses are run to the weak scale. In Section~\ref{soft} we will see that this can make it harder to obtain viable spectra with large splitting between different generation sfermion masses. Additionally, one might be legitimately concerned about whether the \kahler contribution will dominate over other generic contributions that may be expected to also couple the SUSY breaking and visible sectors with suppression by the string scale. If the two sectors are approximately sequestered, with communication only occurring through the U(1) gauge multiplet and messenger fields, the only extra contribution will be a small, generation universal, anomaly mediated soft mass. This is the scenario we study in detail in Section~\ref{soft}, by taking the parameter $\lambda=1$. However, the extent to which two sectors may be completely sequestered is still unclear (for example, see \cite{Kachru:2007xp,Anisimov:2001zz,DeWolfe:2002nn}). Alternatively $\lambda$ can be fairly small $\sim 0.01$, slightly lowering the scale of mediation. This will enhance the \kahler and gauge mediated contributions sufficiently that they can dominate over couplings suppressed by the string or Planck scale.\footnote{In full string constructions it can sometimes be the case that the \kahler contribution is only one of a number of similar sized universal contributions (at least between the first two generations)\cite{Dudas:2008qf,Choi:2006bh}. We do not consider such modifications here.}

%%%%%%%%%%%%%%%%%%%%%%%%%%%%%%%%%%%%%%%%%%%%%%%%%%%%%%%%%%%%%%%%%%%%%
%%%%%%%%%%%%%%%%%%%%%%%%%%%%%%%%%%%%%%%%%%%%%%%%%%%%%%%%%%%%%%%%%%%%%

\section{MSSM Spectra} \label{soft}
Having discussed the main features of our models, in this Section we study in some detail the pattern of soft terms obtained in the MSSM sector. The spectra of soft masses in the previous sections are valid at the energy scale where SUSY breaking is mediated to the visible sector. For the gauge and \kahler contributions this is the mass of the messenger fields and the SUSY breaking sector respectively.

To make any phenomenological predictions it is necessary to run the soft masses to the weak scale. While doing this there will be two dominant and
competing effects on the stop masses \cite{ArkaniHamed:1997ab}: 1) The non-zero gaugino masses will tend to pull the third generation
soft masses squared to larger values, as in gaugino mediated scenarios, and 2) The large first and second generation masses from \kahler mediation will push the third generation
soft masses squared towards negative values.  In cases of very low scale mediation these effects have a reasonably small impact due to the small size of the logs, while in models with higher scale mediation these effects can be significant and limiting of the low energy spectra that can be obtained.  First we consider the field theory case, with low scale SUSY breaking, using the particular example of the Polonyi Model discussed in Section~\ref{pol}, then we examine the sting motivated case with the ISS model of Section~\ref{softiss}.

\subsection{Polonyi Model} \label{softpol}
Recall, $F_{\rm{eff}}= \epsilon^6 v_{}^2$, $m_{\rm{mess}}= \epsilon^3 v_{}$, and a \kahler mass contribution $m_K= \epsilon^4 v_{}$. The charge assignment to MSSM fields is given by Table \ref{Tab1}, therefore the Standard Model fermion masses dictate $\epsilon \sim 0.1$, and hence to obtain a reasonable spectrum of soft terms requires  $M_*\sim 10^8 \GeV$ and $v_{} \sim 10^{7}\GeV$. 

\begin{table}[h]
\noindent \centering{}%
\begin{tabular}{|c|c|c|c|c|c|}
\hline 
 & $q_{L}$ & $u^{c}$ & $e^{c}$ & $L$ & $d^{c}$\tabularnewline
\hline 
\hline 
generation 1 & $1$ & $1$ & $1$ & $1$ & $1$\tabularnewline
\hline 
generation 2 & $1$ & $1$ & $1$ & $1$ & $1$\tabularnewline
\hline 
generation 3 & $0$ & $0$ & $0$ & $0$ & $0$\tabularnewline
\hline 
\end{tabular} ~~~%
\begin{tabular}{|c|c|c|c|c|}
\hline 
$H_{u}$ & $H_{d}$ \tabularnewline
\hline 
\hline 
$0$ & $0$ \tabularnewline
\hline 
\end{tabular} \caption{Charge assignments for low scale breaking}
\label{Tab1}
\end{table}

As discussed, the third generation superfields are uncharged and hence obtain Yukawas of $\mathcal{O}\l(1\r)$, while mass terms for the first two generations have non-zero net $U(1)_{}$
charge therefore are generated only once $S_{-1}$ gains a VEV.  Due to the GUT-consistent structure of charges, the lepton mass hierachy is parametrically the same as
that of the down-type quarks, although the two sets of coefficients are not equal. The resulting up- and down-like Yukawas
are given by
\begin{equation}
U = \left(\begin{array}{ccc}
c_{11} \epsilon^2 & c_{12} \epsilon^2 & c_{13}\epsilon\\
c_{21} \epsilon^2 & c_{22} \epsilon^2 & c_{23} \epsilon\\
c_{31} \epsilon & c_{32} \epsilon & c_{33}
\end{array}\right), \qquad  D=  \left(\begin{array}{ccc}
c'_{11} \epsilon^2 & c'_{12} \epsilon^2 & c'_{13} \epsilon\\
c'_{21} \epsilon^2 & c'_{22}  \epsilon^2 & c'_{23} \epsilon\\
c'_{31} \epsilon & c'_{32}  \epsilon & c'_{33}
\end{array}\right),
\end{equation}
where $c_{ij}$ and $c'_{ij}$ are coefficients which, as discussed, are not subject to any symmetry structure.
Before inclusion of these coefficients the U(1) charges lead to a mass spectrum of SM fermions parametrically of the form
\begin{equation}
m_{up}  \sim  \left\langle H_{u}\right\rangle \left(\begin{array}{ccc}
1 & \epsilon^2 & \epsilon^2\end{array}\right)\qquad
m_{down}  \sim  \left\langle H_{d}\right\rangle \left(\begin{array}{ccc}
1 & \epsilon^{2} & \epsilon^{2}\end{array}\right) \sim m_{lepton},
\end{equation}
while the 2-3 block of the CKM matrix is of the successful form
\begin{equation}
\label{CKM1}
V_{CKM}\sim\left(\begin{array}{cc}
 1 & \epsilon\\
 \epsilon & 1
\end{array}\right) .
\end{equation}
As discussed in Section~\ref{field} the mixings and mass-hierarchies involving the first generation are not set by the broken U(1) but depend on the coefficients $c_{ij}$ and $c'_{ij}$ for
$i~{\rm or}~j\in \{1,2\}$ which depend upon independent physics.  This physics might be an additional UV flavour symmetry that is independent of SUSY breaking dynamics, or it might
be the result of a random anarchic structure.  For instance, if the $\mathcal{O}(1)$ coefficients $c_{ij}$ and $c'_{ij}$ take random values over a finite range, for example a flat distribution in $\l[0,1\r]$, the total $3\times 3$ CKM structure can easily be close to that observed. Additionally, these coefficients and level repulsion in the eigenvalues of the mass matrices can account for the fairly large splitting observed between the first two generation fermions. In any case, in our model, there is strong alignment between the third generation sfermion and fermion mass eigenstates. Typically, the first two generation fermion mass eigenstates contain at most a component of size $\epsilon$ of the third generation U(1) eigenstate, while the first two generation sfermion masses are equal to high precision. 

%Since the down sector masses are suppressed by a factor
%of $\epsilon$ , $\tan\beta=\frac{\left\langle H_{u}\right\rangle }{\left\langle H_{d}\right\rangle }\sim\frac{m_{t}}{m_{b}}\epsilon\sim5$
%assuming Yukawa coefficients in the third generation are $\mathcal{O}(1)$.
%This may be phenomenologically favoured over the alternative of very
%large $\tan\beta\sim\frac{m_{t}}{m_{b}}\sim40$ in enhancing the Higgs mass to $125\GeV$ in an NMSSM like model (see \cite{Ellwanger:2009dp} for a review). As the bottom Yukawas
%are small, even though we will ultimately find multi-TeV scale bottom squarks these do not lead to fine tuning of the electroweak scale.

In order to study the spectrum of sfermion masses that may occur in such a theory it is most interesting to fix the gauge mediated contribution
to these masses so that the gluino is $\sim 1.5~\TeV$ close to current limits. This fixes the combination
\begin{equation}
m_{gauge}\sim\l(\frac{\alpha_3}{4\pi}\r) \epsilon^3 v_{} \sim 10^3 \GeV .
\end{equation}
Therefore the \kahler contribution is given by
\begin{equation} 
m_K \sim \epsilon^4 v \sim \l(\frac{4\pi}{\alpha_3}\r) \epsilon m_{gauge} \label{polkah} ,
\end{equation}
which depends only on the value of the parameter $\epsilon$. In addition, we choose the number of pairs of messenger fields $n_m=5$. This increases the gauge mediated gluino mass, which is proportional to $n_{m}$, relative to the stop mass which is proportional to $\sqrt{n_m}$, but is not so large as to lead to a Landau pole for the SM gauge couplings
below the GUT scale. \footnote{In this model, anomaly cancellation requires additional matter charged under the U(1) and MSSM gauge groups. We assume these fields have charges such that they do not couple strongly to the SUSY breaking sector, and are not sufficiently numerous that they lead to a Landau pole. Alternatively, anomaly cancellation with no extra matter is possible if there are fewer messenger fields present. The only effect of such a modification is the gluino mass will be lowered towards that of the stop.} 

\begin{figure}
\centering
\includegraphics[scale=0.7]{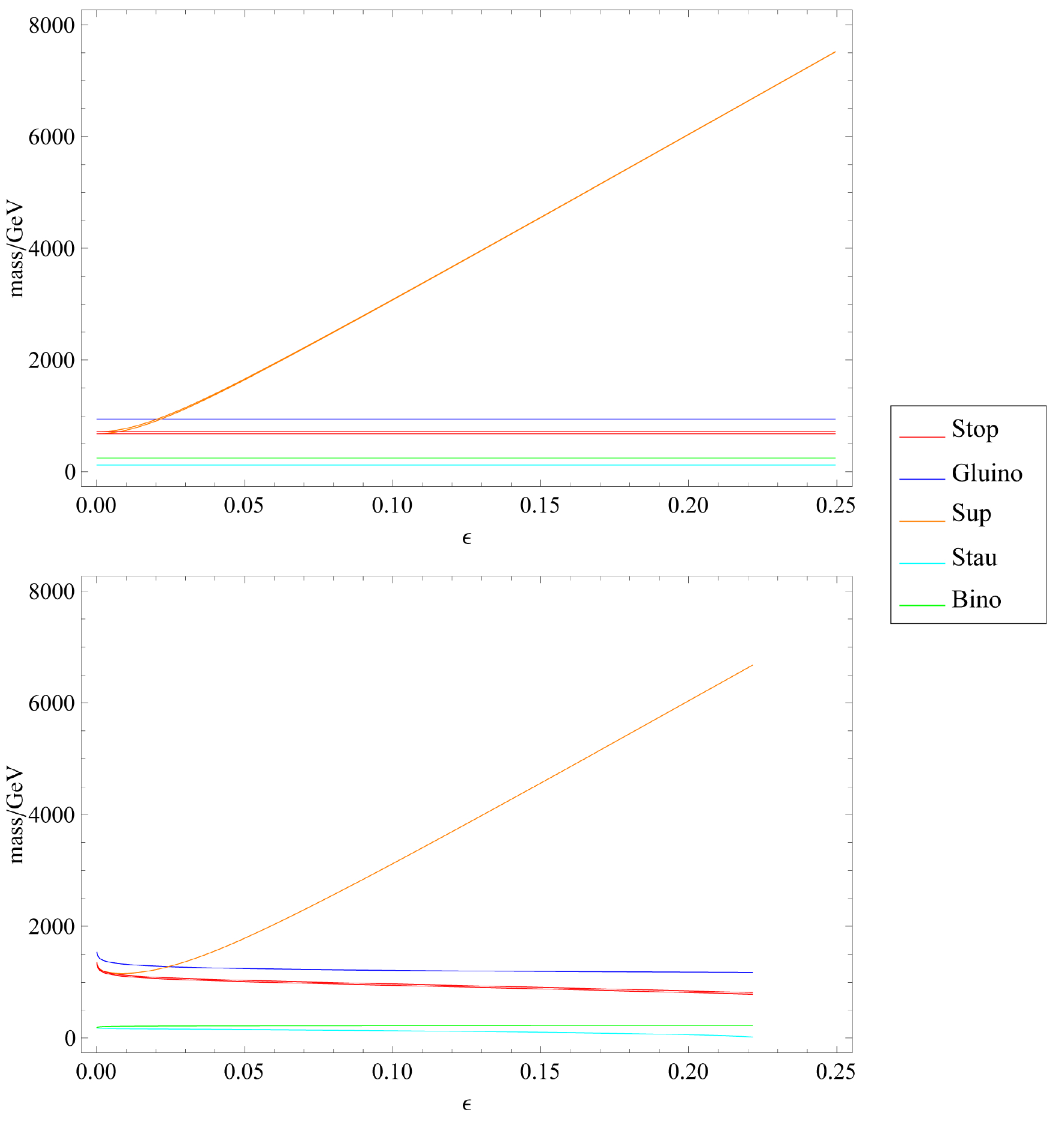} 
\caption{The spectrum of superparticles before (top) and after (bottom) running to the weak scale in the Polonyi model. $F$ and $v$ have been fixed to give a gluino in the region of $1.3\TeV$ after running, close to current LHC limits, while
$M_{*}$ is varied changing $\epsilon$ and therefore the relative
importance of the \kahler interactions. For $\epsilon > 0.22$ the first two generation sfermions are so heavy that a stau is driven tachyonic during running and the weak scale spectrum is not phenomenologically viable.} 
\label{Fig1} 
\end{figure}

In Fig.\ref{Fig1} top we plot the soft masses obtained at the scale $\sqrt{F}$ by allowing $\epsilon$ to vary while keeping the gauge mediated contribution fixed. As $\epsilon$ increases the first two generations obtain increasing masses from the \kahler operator resulting in a natural SUSY spectrum. As discussed we need to run the spectrum to the weak scale. The \kahler contribution to the first two generation soft masses turns on at a scale $\sqrt{F}\sim \epsilon^2 v$ while gauge mediated contributions to these and the gaugino and third generation soft masses begins at $m_{mess}\sim \epsilon^3 v$. Depending on the charge assignments, and the particular value of $\epsilon$, it is possible that the sbottom or stop may be driven tachyonic at some point in this energy regime. Such an event is not necessarily problematic if these states run back to positive mass squared before the weak scale. Provided $m_{\tilde{t}}\left(M_Z\right) > \frac{1}{10} M_3\left(M_Z\right)$ the EW breaking vacuum is sufficiently meta-stable against decays to a colour breaking vacuum compared to the lifetime of the universe  \cite{Riotto:1995am,Kusenko:1996jn,Dermisek:2006ey}. This relation is typically satisfied for our models.\footnote{The energy region where such states are tachyonic is fairly small hence there is little danger of reheating after inflation into a colour breaking vacua, and even in this case it has been suggested that the EW vacua may be favoured \cite{Riotto:1996xd}.}

Below a scale $m_{1,2}$ the first two generation sfermions are integrated out of the theory and have no further effect on the third generation running, while the positive contribution from the gluino persists until the gluino mass is reached. Additionally the gauginos and first two generation sfermion masses also flow. We solve the renormalisation group equations numerically and plot the mass spectrum at the weak scale in Fig.\ref{Fig1} bottom panel. As $\epsilon$ increases the \kahler mass contribution increases and during running the stop and stau masses are driven smaller, until at $\epsilon \sim 0.2$ the right handed stau is tachyonic at the weak scale and the spectrum is not phenomenologically viable.

% Similarly, the third generation right handed stau (which does not gain a large \kahler soft mass with the charge assignments ()) renormalisation is dominated by
%\begin{equation}
%\frac{d}{dt}m_{\tilde{e_3}}^2 = -8 \frac{\alpha_1}{4\pi} Y_{e_3} M_1^2 + \frac{32}{5} Y_{e_3} \frac{\alpha_1\alpha_3}{16\pi^2} \tilde{m}_{1,2}^2 ,
%\end{equation}
%where there is a mixed $\alpha_3 \alpha_1$ contribution since all three MSSM copies of the the 5 of SU(5) gain a large \kahler mass whereas only two of the generations of 10s of SU(5) do so. The gauginos and first two generation sfermion masses will also flow. We solve the renormalisation group equations numerically and plot the mass spectrum at the weak scale in Fig.\ref{Fig1} bottom panel. As $\epsilon$ increases the \kahler mass contribution increases and during running the stop and stau masses are driven smaller, until at $\epsilon \sim 0.15$ the stau is driven tachyonic and the spectrum is not phenomenologically viable.

\begin{figure}
\centering
    \includegraphics[trim = 0mm 0mm 0mm 0mm, clip, width
      = 0.45\textwidth]{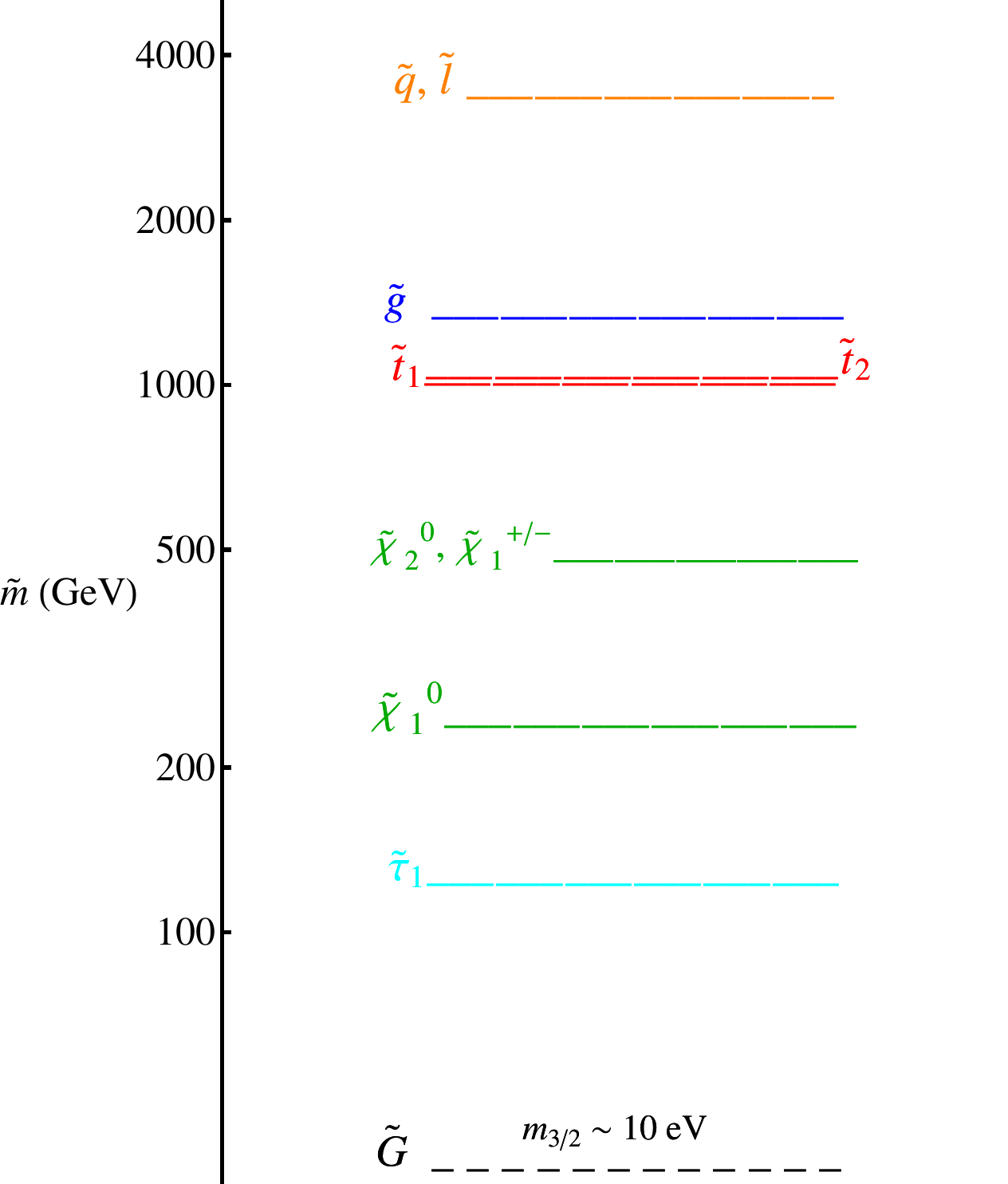}% l b r t -
\caption{A typical spectrum of superparticles after running to the weak scale in the Polonyi model. $F$ and $v$ are fixed to give a gluino in the region of $1.3\TeV$ and as a representative example $\epsilon=0.1$. There is a very light gravitino LSP and the right handed stau is the NLSP.} 
 \label{Fig:spectrum}
\end{figure}

The key point of our models is that for values of $\epsilon$ motivated by the fermion mass heirachy the split between the first two generation soft masses and the third is sufficiently large to realise natural SUSY, but not so large as so lead to tachyonic third generation states. The NLSP (after the gravitino) is typically a stau, which is fairly light, and can modify cosmology and certain collier signals as we will discus later. As a representative example of the full spectra that may typically be obtained, we show the field content for $\epsilon= 0.10$ under the current assumptions in Fig.\ref{Fig:spectrum}. This is a reasonable value in the middle of the plausible range without fine tuning to the edge of the allowed region.

%XXXX
%We can use this to estimate if the stops remain non-tachyonic during running. Neglecting the second order effect of the gauge coupling running (since the running is only over a small energy range) this is easily integrated to
%\begin{equation}
%m_{\tilde{q_3}}^2 \left(100\GeV \right) = m_{\tilde{q_3}}^2 \left(\sqrt{F}\right) + 8 \alpha_3 C_3 M_3^2 \log\frac{\sqrt{F}}{M_3}-32 C_3 %\alpha_3^2 \tilde{m}_{1,2}^2  \log\frac{\sqrt{F}}{\tilde{m}_{1,2}} .
%\end{equation}
%Including what we know about the spectrum this gives
%\begin{equation}
%m_{\tilde{q_3}}^2 \left(100\GeV \right) \sim n_m \l(\frac{\alpha_3}{4\pi}\r)^2 \epsilon^6 v^2  + 8 \alpha_3 C_3 \l(n_m \l(\frac{\alpha_3}%{4\pi}\r) \epsilon^3 v_{}\r)^2  \log\frac{4\pi}{\alpha_3}-32 C_3 \alpha_3^2 \epsilon^8 v^2  \log\frac{1}{\epsilon} ,
%\end{equation}
%assuming we are in the dangerous regime where where the \kahler contribution to the first two generation sfermion masses dominates over the %gauge mediated contribution. It is straightforward to check that this results in stop mass which is always non-tachyonic, essentially the %heavier the first two generations the less time they are active in the renormalisation.

\subsection{ISS Model}
It is also interesting to see to what extent we can realise natural SUSY in the ISS model at a relatively high scale. In this case we take the string motivated \kahler contribution, $m_K\sim \frac{F}{M_*}$. To simplify the analysis we assume the parameter  $\lambda=1$, and take the U(1) charge assignment of Table \ref{Tab2}.\footnote{In this case we are choosing a U(1) charge structure that is not compatible with a traditional 4D GUT.  However it is compatible with an orbifold GUT structure, which can result from an underlying IIB D-brane model, with split matter multiplets \cite{Hall:2001pg,Hebecker:2001wq}. Thus precision SUSY gauge-coupling unification can be maintained.}  This has the phenomenological benefit that it gives the right handed stau a large mass, preventing it running tachyonic, which would otherwise place the strongest limit on the allowed values of $\epsilon$.
\begin{table}[h] 
\noindent \centering{}%
\begin{tabular}{|c|c|c|c|c|c|} 
\hline 
 & $q_{L}$ & $u^{c}$ & $e^{c}$ & $L$ & $d^{c}$\tabularnewline
\hline 
\hline 
generation 1 & $1$ & $1$ & $1$ & $1$ & $1$\tabularnewline
\hline 
generation 2 & $1 $ & $1$ & $1$ & $1$ & $1$\tabularnewline
\hline 
generation 3 & $0$ & $0$ & $1/2$ & $1/2$ & $1$\tabularnewline
\hline 
\end{tabular} ~~~% 
\begin{tabular}{|c|c|c|c|c|}
\hline 
$H_{u}$ & $H_{d}$ \tabularnewline
\hline 
\hline 
$0$ & $0$ \tabularnewline
\hline 
\end{tabular} \caption{Charge assignments for high scale breaking}
 \label{Tab2}
\end{table}
Before inclusion of the $c_{ij}$ and $c'_{ij}$ coefficients these give a mass pattern 
\begin{equation}
m_{up} \sim  \left\langle H_{u}\right\rangle \left(\begin{array}{ccc}
1 & \epsilon^2 & \epsilon^2\end{array}\right)\qquad
m_{down}  \sim \left\langle H_{d}\right\rangle \left(\begin{array}{ccc}
\epsilon & \epsilon^2 & \epsilon^2 \end{array}\right) \sim m_{lepton},
\end{equation}
while the third-second generation sub-block of the CKM matrix is again of the form (\ref{CKM1}).

Reasonable splitting of the third generation leads to $0.007\lesssim\epsilon\lesssim0.05$. This also gives a CKM matrix of the correct form to leading order. Since the third generation down sector masses are suppressed by a factor of $\epsilon$, $\tan\beta=\frac{\left\langle H_{u}\right\rangle }{\left\langle H_{d}\right\rangle }\sim\frac{m_{t}}{m_{b}}\epsilon\sim1$
assuming Yukawa coefficients in the third generation are $\mathcal{O}(1)$. This may be phenomenologically favoured over the alternative of
large $\tan\beta\sim\frac{m_{t}}{m_{b}}\sim40$ in enhancing the Higgs mass to $125\GeV$ in an NMSSM like model (see \cite{Ellwanger:2009dp} for a  review). As the bottom Yukawas
are small, even though the \kahler mass contribution will lead to multi-TeV scale bottom squarks these do not lead to fine tuning of the electroweak scale.

Again we take there to be five pairs of messenger fields, and as well motivated by string compactifications, $M_* = 10^{16} \GeV$. In order to obtain a gauge mediated contribution to soft masses (and in particular the gaugino masses) of order $\TeV$ such that these are close to current limits but not excluded requires $\Lambda \sim 10^{10}\GeV$. To obtain  \kahler contributions to the first two generation masses that are also a few TeV for reasonable values of $\epsilon$, we take the charge, introduced in Section~\ref{softiss}, $n=4$. Of course, this is a particular choice which leads to viable natural spectra, however as a discreet value it is plausible and not a fine tuning in the sense of a continuous parameter. Such a choice also has the benefit of setting the mass of the messengers and also the coefficient of $Q_i\tilde{Q}^i$ in the electric ISS superpotential equal to $\sim \epsilon^3 v_{} \sim 10^{8}\GeV$. Since this is much less than the strong coupling scale of the ISS theory it is valid to use the Seiberg dual of this theory, and the SUSY breaking vacua obtained is sufficiently long lived.

\begin{figure}
\centering
\includegraphics[scale=0.7]{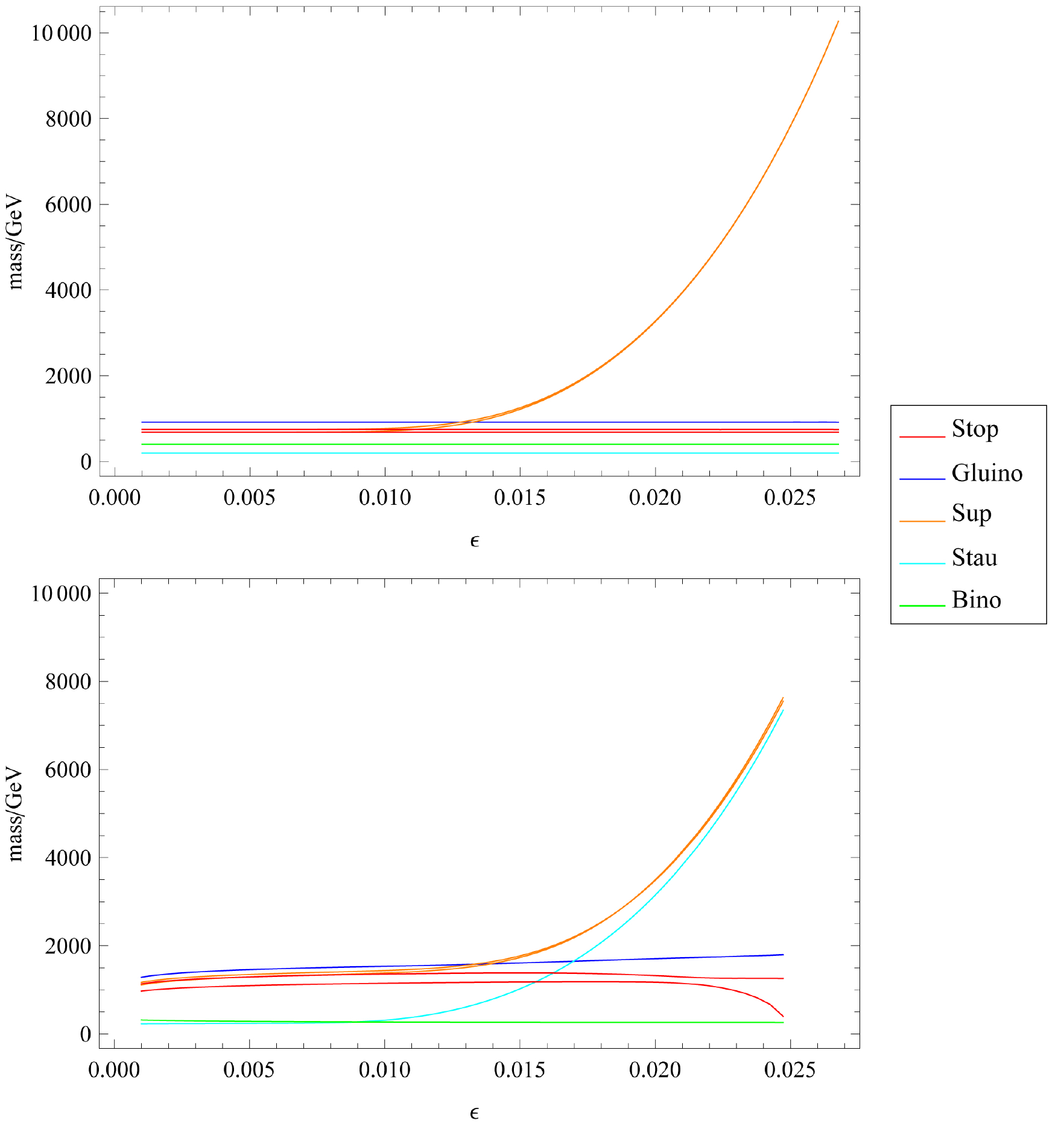} 
\caption{The spectrum of superparticles before (top) and after (bottom) running to the weak scale in the string motivated case. $\Lambda$ and $M_*$ have been fixed to give a gluino in the region of $1\TeV$ after running, and $\epsilon$ is allowed to vary.} 
\label{Fig2} 
\end{figure}

Unlike the field theory case, the gauge mediated contribution, $m_{\rm{gauge}} \sim \l(\frac{\alpha_3}{4\pi}\r) \frac{\Lambda^2}{M_*}$, is independent of $\epsilon$. Therefore in studying the spectra we simply fix $\Lambda$ and $M_*$ and allow $\epsilon$ to vary, which changes the \kahler mediated contribution $m_K \sim \epsilon^4 \Lambda$.  Since the running occurs over a long period it is important to use the full renormalisation group equations and our analysis is done using {\tt SOFTSUSY} \cite{Allanach:2001kg}. The spectrum obtained before and after running is shown in Fig.\ref{Fig2}.

For choices of $\epsilon \simeq 0.02$ a natural spectrum with light stops and heavy first two generations is obtained, however the range of $\epsilon$ that generates such a spectrum is smaller than in the field theory case. There are two reasons for this, firstly, since SUSY is broken at a higher scale there is more running, therefore for a given stop mass the first two generations cannot be as heavy as previously. This effect is unavoidable in any model of natural SUSY that uses high scales as may be natural in string completions. Secondly, in this scenario the \kahler masses have a power law dependence on $\epsilon$ while the gauge mediated contribution has no such dependence, therefore relatively small changes in $\epsilon$ change the \kahler masses significantly. This may be regarded as a defect of the model, however our purpose is only to demonstrate that natural SUSY spectra are possible in realistic completions.

As previously discussed, in order to obtain a vanishing cosmological constant, there must be an R-symmetry breaking constant superpotential term generated by the theory. Depending on the particular dynamics, the sector that generates the VEV for the S fields may play such a role.

\section{Variations and Signatures} \label{pheno}

\subsection{Variant Spectra}

So far we have not addressed the $\mu/B_\mu$ problem that is very commonly found in models of gauge mediated SUSY breaking \cite{Dvali:1996cu}. This may be solved using a mechanism completely separate from the U(1) and generation of a natural SUSY spectrum, or alternatively, with minor alternations to the charge structure could be solved automatically in our models.  Suppose the down type Higgs has charge $\frac{1}{2}$ under the U(1) symmetry. Then suitable choices of the charges of the lepton and down type superfields can still lead to viable fermion mass patterns and natural soft mass spectra (for example one may shift the charge assignments of all three generations of $d^c$ and $L$ fields by $-\frac{1}{2}$ from their values in Table \ref{Tab2}). As a consequence both the $\mu$ and $B_\mu$ terms are forbidden at tree level, while the down type Higgs obtains a significant \kahler soft mass $m_{H_d}^2\sim \left(4\TeV\right)^2$ while $m_{H_u}^2 \sim \left(200\GeV\right)^2$. If the DSB and messenger sectors additionally involve fields with charge $\pm \frac{1}{2}$ gaining VEVs or F-terms, it is possible to generate $\mu$ and $B_\mu$, and depending on the explicit model, these satisfy the standard relation $B_\mu \sim 16\pi^2 \mu^2$. However, this pattern of soft masses and parameters with $m_{H_d}^2$ large now realises lopsided gauge mediation \cite{DeSimone:2011va}, at least as far as the Higgs sector of the theory is concerned, and which leads to viable EW symmetry breaking without excessive fine tuning.

While we have studied charge assignments such that the first two generation sfermions gain large soft masses, there is an alternative option that can lead to natural spectra. If the top superfields are charged appropriately, it is possible the stops gain a \emph{negative} mass contribution from the \kahler couplings pushing them to a lower mass than the first two generation sfermion masses. Such charge assignments can also generate the fermion mass heirachy if the Higgs multiplets are charged under the U(1). In fact, even if the theory is such that the stops have negative mass squared at the mediation scale it is possible,
after running, to obtain viable spectrum with non-tachyonic stops at the weak scale if the gluino is sufficiently heavy. This was first raised as a possibility in \cite{Dermisek:2006qj} where it was suggested as a mechanism for obtaining a spectrum with low fine tuning. In such a model the stop may be expected to be tachyonic for a relatively large range of energies. Hence, even if the lifetime of the EW breaking and colour preserving vacuum is sufficiently long, there is a concern about whether the universe is likely to find itself in this metastable state after reheating. While we have not investigated this scenario in detail, preliminary investigation demonstrates that it is possible to obtain reasonably natural spectra consistent with LHC constraints. However, as in the models we have focussed on, very light stops, possibly down to $\sim 400 \GeV$, require some fine tuning of the parameters of the theory. 

As an additional possibility, if we reject the requirement of naturalness, it is also easily possible to generate a split \cite{Giudice:2004tc,ArkaniHamed:2004yi} or mini-split spectrum within these models \cite{Arvanitaki:2012ps}. This could occur if all the quark superfields have the same sign charge under the U(1), hence all sfermions gain a large positive mass contribution from \kahler interactions. In order to obtain a viable fermion mass spectrum this would require the Higgs fields to have the opposite charge. In this case since the SUSY flavour problem is solved by decoupling, the U(1) could also generate the texture in the first two generation fermion masses.

\subsection{Collider Signals, Flavour and Higgs} \label{coll}

The collider signals of the natural spectra typical of our models have been studied
extensively. Depending on the charge assignments, a bino or stau is generically the NLSP, which for very
low gravitino mass may decay in a typical detector distance while for larger
gravitino mass will escape the detector. Both cases lead to clear signals
that can be studied at the LHC. However, the relatively heavy gluino
masses ($m_{\tilde{g}}> 1.5{\rm TeV}$) and especially the almost decoupled first two generations, reduce production cross
sections dramatically, and spectra are typically well within current
LHC limits such that a light stop is not ruled out. As we have seen, a very light stop is hard to achieve, a more realistic model has been seen in Fig.\ref{Fig:spectrum}. Since in this case the stop is not especially light (though still far below current bounds for squark masses in generation universal models) such a spectrum will be challenging to discover at the LHC until a large integrated luminosity has been accumulated. Additionally, if R-parity is broken, spectra with a lighter gluino may be compatible with LHC constraints, allowing lighter stops in our models \cite{Allanach:2012vj}. More detailed analysis of the expected signals and phenomenology can be found in for example \cite{Lee:2012sy,Espinosa:2012in,Larsen:2012rq,Han:2012fw,Drees:2012dd,Brust:2011tb,Meade:2006dw,Barbieri:2010ar}.

A common concern in SUSY theories is suppressing flavour changing effects to safe levels. In our model these effects can be well within current limits. The ordinary gauge mediated contribution is automatically flavour blind as normal. Additionally the \kahler contribution to the first two generation sfermions is universal. Therefore flavour changing effects occur only due to the small mixing in the CKM matrix between the first two generations and the third generation. More precisely, the first two fermion mass eigenstates include only a component of the 3rd generation U(1) eigenstate of size $\epsilon$. In order to produce a realistic CKM matrix $\epsilon$ must satisfy $\epsilon\sim V_{cb}$. Hence, the sfermion mass squared matrix differs from diagonal in the first two generation sector at most by by elements like $V_{cb}^2 m_{\tilde{t}}^2$. There is also additional suppression of flavour changing effects due to the relatively large masses of the first two generation sfermions. Utilising the expressions in \cite{Gabbiani:1996hi} we find that CP conserving flavour changing effects are typically well within experimental limits. CP violating processes generally give stronger constraints; if the first two generation sfermions are near their maximum allowed mass and $\epsilon$ fairly small these can be within current limits for $\mathcal{O}\left(1\right)$ phases in the soft terms. Alternatively, we can assume the UV theory is such that these phases are small or zero. \footnote{The counting of physical phases also depends on the mechanism that generates $\mu$ and $B\mu$, hence is model dependent.}

While light stops allow a theory without excessive fine tuning of the electroweak scale, it is not immediately obvious how to combine such a spectrum with a lightest Higgs mass of $126 \GeV$ as recently discovered by ATLAS and CMS. The issue arises since, in the MSSM, the Higgs mass is bounded by $M_Z$ at tree level. For the motivated spectra found in the previous sector it is hard to obtain a sufficiently large Higgs mass through stop loop corrections while keeping fine tuning low. A better alternative is to implement an NMSSM like model or one of the related extensions of the MSSM. Such models are may be independently motivated to generate an appropriately sized $\mu$ parameter, and give an additional source of the quartic Higgs coupling raising the Higgs mass to the required value. Low fine tuning may even prefer very large values of the parameter $\lambda$ \cite{Hall:2011aa} that may run non-perturbative at an intermediate scale which could be identified as $M_*$ in our theories while still not destroying the success of SUSY gauge coupling unification \cite{Hardy:2012ef}.
Finally, in the case of charge assignments that lead to a light stau (for example if we impose a GUT compatible structure), the Higgs to two photon decay rate may be enhanced which could account for the very tentative hints of an excess over SM rates at the LHC \cite{Carena:2012gp}. Alternatively in a $\lambda$SUSY-like model this rate may be enhanced by light charginos \cite{SchmidtHoberg:2012yy}.

\subsection{Axions and Cosmology}
%%%%%%%%%%%%%%%%%%%%%%%%%%%%%%%%%%%%%%%%%%%%%%%%%%%%%

In the string motivated case an approximate global symmetry is spontaneously broken and there will be an axion present in the low energy theory, which will, because
of the $U(1)\times G_{MSSM}^{2}$ anomaly, have couplings to the MSSM gauge multiplets. Therefore in the case of relatively high scale mediation this state could even play
the role of the QCD axion. This is by no means necessary, however.  For example, there may be couplings between the axion and any hidden gauge groups, for instance
in the DSB sector, depending on the particular anomaly coefficients of the theory.  Such anomalous couplings to a hidden gauge sector typically imply that the
axion-like states gains a large mass of order $\frac{\Lambda_{\rm{hidden}}^2}{f_a}$, and therefore cannot be the QCD axion. On the other hand this allows current astrophysical
and direct search bounds to be easily evaded even if $f_a\sim v\ll 10^9\GeV$. 

Further, depending on the mass and decay constant of the axion, as well as the initial misalignment angle and thermal history of the universe, this can provide a significant component of the dark matter. In fact it may be highly beneficial to couple the QCD axion to the DSB sector: typically overproduction of the axino and saxion, combined with gravitino limits, strongly constrains the reheat temperature over a large parameter space \cite{Cheung:2011mg}. However if there is a significant coupling between the axion multiplet and the SUSY breaking sector the axino and saxion can gain large masses greatly relaxing these limits \cite{Higaki:2011bz}. The presence of a light axion degree of freedom coupling both to the DSB and visible sectors is similar to a scenario we recently studied where the axion was the primary mediator of SUSY breaking \cite{Baryakhtar:2013wy}, although in the present models the axion multiplet does not typically gain a significant F-term.

%%%%%%%%%%%%%%%%%%%%%%%%%%%%%%%%%%%%%%%%%%%%%%%%%%%%%

Apart form the possibility of such an axion, the cosmology of the models are fairly similar to that of normal gauge
mediated models. One exception is when the U(1) charge assignments are such that there is a light stau in the theory, in which case it is typically the NSLP after running (see for example Fig.\ref{Fig:spectrum}). This may be beneficial for cosmology; since a stau NLSP leads to decay hadronically it can decay into the gravitino later than other NLSP candidates without disrupting BBN. As a result a heavier gravitino is compatible with observations, permitting a higher reheat temperature without gravitino overproduction aiding inflation model building. More precisely, it has been suggested that $F$ in the region $\sqrt{F}\sim10^{8.5\div10}\GeV$ and thus $m_{3/2}\sim 0.1 \div 100\GeV$ (which is compatible with a GUT scale value of $M_*$) may permit reheat temperatures up to $\sim10^{9}\GeV$ \cite{Asaka:2000zh}.

\section*{Acknowledgements}

\noindent We are grateful to Asimina Arvanitaki, Nathaniel Craig, Savas Dimopoulos, Xinlu Huang, Ken Van Tilburg, James Unwin, Giovanni Villadoro and
especially Masha Baryakhtar for useful discussions. This work was supported in part by ERC grant BSMOXFORD no. 228169. JMR thanks the
Stanford Institute for Theoretical Physics for their hospitality during the early and late stages of this
work. The authors also thank the CERN Theory Group for their hospitality. EH thanks Merton College Oxford for a travel grant.

%%%%%%%%%%%%%%%%%%%%%%%%%%%%%%%%%%%%%%%%%%%%%%%%%%%%%%%%%%%%%%%%%%%%%
%%%%%%%%%%%%%%%%%%%%%%%%%%%%%%%%%%%%%%%%%%%%%%%%%%%%%%%%%%%%%%%%%%%%%
%%%%%%%%%%%%%%%%%%%%%%%%%%%%%%%%%%%%%%%%%%%%%%%%%%%%%%%%%%%%%%%%%%%%%

\appendix
\section{Source of the contact operator} \label{apenA}

In this Appendix we provide a justification for the interactions \eqref{kah} \eqref{kahiss} that are crucial to our model.  In the pure field theory case this is obtained
by integrating out the heavy gauge multiplet, as discussed in \cite{ArkaniHamed:1998nu} directly leading to an effective term in the \kahler potential
\begin{equation}
\mathcal{L} \supset  -\sum_{i,j}\frac{g^2 q_i q_j }{M_{Z'}^2}  \int d^4\theta\,\,  \phi^{\dagger i} \phi_i \phi^{\dagger j} \phi_j
\end{equation}
where $\phi_i$ and $\phi_j$ are any fields charged under the gauge symmetry. Alternatively this can be regarded as the vector multiplet gaining a D-term.
In the Stueckelberg case, the interaction can also be understood in this way, however it is interesting to also understand it directly from the Lagrangian. For a vector multiplet $V$, which gains a mass $M$ through interaction with a Stueckelberg field $S$, and is coupled to hidden sector fields $\Phi$ and MSSM fields $Q$ this is given by:
\begin{equation}
\int d^{4}\theta\left(\Phi^{\dagger}e^{q_{\Phi}gV}\Phi+Q^{\dagger}e^{q_{Q}gV}Q + M^2 \left(V + \frac{1}{M}\left(S-S^\dagger \right) \right)^2 \right) .
\end{equation}
In the limit that $M$ is much greater than any other mass scale in the theory, $V$ may be integrated out by solving $\frac{\partial K}{\partial V}=0$ with solution
$V = -(g/2 M^2) \left( q_{\Phi} \Phi^\dagger \Phi + q_{Q} Q^\dagger Q + (S-S^\dagger)/g M \right)$. Inserting this back into the original \kahler potential leads to
\begin{equation}
\int d^{4}\theta \left(- \frac{g^2 q_\Phi q_Q}{M^2}  \Phi^\dagger \Phi Q^\dagger Q\right) .
\end{equation}
The bilinear dependence of these terms on the charges justifies the claim in the text that for a suitable charge assignment a positive mass$^2$ contribution to sfermion masses 
arises and also that a sign difference in the charge of $Q_3$ relative to $Q_{1,2}$ can lead to mass terms of the opposite sign.
In the string context we expect $M\sim gM_*$ which results in the dependence on the gauge coupling dropping out.\footnote{The limit $g\to 0$ is obscured by these terms as the mass scale $M\sim gM_*$ is no-longer large.  In the leading and sub-leading terms we have dropped a overall model-dependent coefficients.}
Finally, higher terms in the expansion occur, e.g.,
\begin{equation}
\int d^{4}\theta \left( \frac{g^2 q_\Phi^2 q_Q}{M^4} \Phi^\dagger  \Phi^\dagger \Phi \Phi Q^\dagger Q \right) .
\end{equation}
Such terms, however, are harmless in the relevant parameter range for our discussion.

\section{Gauge mediated contribution to soft masses} \label{apenB}

In the main body of the text an approximate expression for the gauge-mediated contribution to soft masses was quoted.
Here we give the precise formulae as used in our numerical studies and figures.

The two contributions,
\eqref{gaugecoupling} and \eqref{kah} in the Polonyi case, and \eqref{gaugecouplingiss} and \eqref{kahiss} in the ISS case, 
result in a messenger scalar mass matrix (assuming $\Psi$ and $\Psi^{c}$ have the same $U(1)_{_{}}$ charge; the analysis
is straightforwardly extended to other cases) of the form
\begin{equation}
\mathcal{L}\supset\left(\begin{array}{cc}
\tilde{\Psi}^{\dagger} & \tilde{\Psi^{c}}^{\dagger}\end{array}\right)\left(\begin{array}{cc}
m_{mess}^2 + m_K^2 & F_{eff}\\
F_{eff} & m_{mess}^2 + m_K^2 
\end{array}\right)\left(\begin{array}{c}
\tilde{\Psi}\\
\tilde{\Psi^{c}} ,
\end{array}\right) \label{messmass}
\end{equation}
where $m_{mess}$ is the supersymmetric mass of the multiplet, $m_K$ is the mass due to the \kahler interactions, and $F_{eff}$ is the effective F-term felt by messengers through the superpotential.

Hence, the messenger scalar mass eigenstates are given by $\frac{1}{\sqrt{2}}\left(\tilde{\Psi}\pm\tilde{\Psi^{c}}\right)$ with masses
$m_{1,2}^{2}  = \left(m_{SUSY}^2+ m_K^2 \pm F_{eff}\right)$, while the messenger fermion masses are simply $m_f = m_{SUSY}$.
Gaugino masses are generated through the normal diagrams of gauge mediation, and are given by
\begin{equation}
m_{\lambda i}=\frac{\alpha_{i}}{4\pi}n_{m}m_{f}\left(\left(\frac{m_{1}^{2}}{m_{1}^{2}-m_{f}^{2}}\right)\log\left(\frac{m_{1}^{2}}{m_{f}^{2}}\right)-\left(\frac{m_{2}^{2}}{m_{2}^{2}-m_{f}^{2}}\right)\log\left(\frac{m_{2}^{2}}{m_{f}^{2}}\right)\right) , \label{gauginomass}
\end{equation} 
which reduces to that commonly found through analytic continuation \cite{ArkaniHamed:1998kj}
in the limit $F_{eff} \gg m_{K}^2$ 
and $F_{eff}\ll m_{SUSY}^2$ . However, for the values of parameters we are interested in, these
conditions are not satisfied and the full expression (\ref{gauginomass}) is required.

The contribution to sfermion masses from gauge mediation with these messenger masses is given by
\begin{equation}
m_{gauge}^{2}=\sum_{i}\left(\frac{\alpha_{i}}{4\pi}\right)^{2}\, C_{i} \, n_{m}\, \bigl[g (m_{1},m_{2},m_{f})+h (m_1,m_2,m_f,\Lambda_{UV}) \bigr] .
\label{gaugemed}
\end{equation}
Here $C_{i}$ is the quadratic Casimir of the scalar and
\begin{equation}
h\left(m_1,m_2,m_f,\Lambda_{UV}\right)= - \left(2m_1^2+2m_2^2 - 4m_f^2\right)\log\left(\frac{\Lambda_{UV}^2}{m_f^2}\right) . 
\end{equation}
while $g(m_{1},m_{2},m_{f})$ is the contribution from the normal diagrams of gauge mediation whose lengthy explicit form is given in \cite{Poppitz:1996xw}. 
As also discussed in \cite{Poppitz:1996xw}, the non-vanishing supertrace of the messenger sector results in a negative contribution, described by $h\left(m_1,m_2,m_f,\Lambda_{UV}\right)$. due to the need to include a counterterm for $\epsilon$-scalar masses in dimensional regularisation. The scale $\Lambda_{UV}$ is the mass at which additional states charged under the MSSM gauge group appear resulting in a vanishing supertrace.  For our theories this is naturally $\Lambda_{UV}=M_*$. Due to their large mass these extra states do not contribute significantly to MSSM masses as messengers of gauge mediation.

%\begin{eqnarray}
%g(m_{1},m_{2},m_{f})=\log^{2}\left(m_f^{2}\right)\left(m_1^{2}+m_2^{2}+2m_f^{2}\right)+\frac{1}{2} [ 6\left(m_1^{2}+m_2^{2}-2m_f^{2}\right)+&&\nonumber\\
%\log^{2}\left(m_1^{2}\right)\left(m_1^{2}-m_2^{2}+2m_f^{2}\right)+\log^{2}\left(m_2^{2}\right)\left(-m_1^{2}+m_2^{2}+2m_f^{2}\right)+&&\nonumber\\
%2\log\left(m_1^{2}\right)\left(\left(m_1^{2}+m_2^{2}\right)\log\left(m_2^{2}\right)+2m_1^{2}\right)+4m_2^{2}\log\left(m_2^{2}\right) ]-&&\nonumber\\
%2\log\left(m_f^{2}\right)\left(\left(m_1^{2}+m_f^{2}\right)\log\left(m_1^{2}\right)+\left(m_2^{2}+m_f^{2}\right)\log\left(m_2^{2}\right)+2m_f^{2}\right)+&&\nonumber\\
%m_1^{2}\left(-\text{Li}_{2}\left(1-\frac{m_1^{2}}{m_2^{2}}\right)\right)-m_2^{2}\text{Li}_{2}\left(1-\frac{m_2^{2}}{m_1^{2}}\right)+&&\nonumber\\
%2\left(m_1^{2}+m_f^{2}\right)\text{Li}_{2}\left(1-\frac{m_1^{2}}{m_f^{2}}\right)+2\left(m_f^{2}-m_1^{2}\right)\text{Li}_{2}\left(1-\frac{m_f^{2}}{m_1^{2}}\right)+&&\nonumber\\
%2\left(m_2^{2}+m_f^{2}\right)\text{Li}_{2}\left(1-\frac{m_2^{2}}{m_f^{2}}\right)+2\left(m_f^{2}-m_2^{2}\right)\text{Li}_{2}\left(1-\frac{m_f^{2}}{m_2^{2}}\right)
%\end{eqnarray}

Unlike gaugino masses, sfermions gain significant masses from gauge mediation even when
the messenger masses are dominated by the diagonal \kahler contribution. The reason for this is clear: the \kahler contribution is effectively a D-term mass and does
not break an R-symmetry, which protects gaugino masses but not sfermion masses.

\bibliography{retro}
\bibliographystyle{JHEP}

\end{document}